\documentclass[twocolumn,letterpaper,aps,10pt,pra]{revtex4-1}

\usepackage[dvips]{graphicx}
\usepackage{amssymb,amsfonts,amsmath}
\usepackage{bbm}
\usepackage{upgreek}

\newcommand{\defeq}{:=}
\newcommand{\ket}[1]{|#1 \rangle}

\usepackage[usenames,dvipsnames]{color}
\definecolor{LinkColor}{rgb}{0,0,.5}
\usepackage[colorlinks=true,linkcolor=red,citecolor=LinkColor,urlcolor=LinkColor]{hyperref}

\begin{document}
\title{Time-resolved magnetic sensing with electronic spins in diamond}
\author{A. Cooper, E. Magesan, H.N.  Yum, P. Cappellaro}
\email{pcappell@mit.edu}
\affiliation{Department of Nuclear Science and Engineering and Research Laboratory of Electronics,\\Massachusetts Institute of Technology, 77 Massachusetts Avenue, Cambridge, Massachusetts 02139, USA}
\begin{abstract}
Quantum probes can measure time-varying fields with high sensitivity and spatial resolution, enabling the study of biological, material, and physical phenomena at the nanometer scale. In particular, nitrogen-vacancy centers in diamond have recently emerged as promising sensors of magnetic and electric fields.
Although coherent control techniques have measured the amplitude of constant or oscillating fields, these techniques are not suitable for measuring time-varying fields with unknown dynamics. Here we introduce a coherent acquisition method to accurately reconstruct the temporal profile of time-varying fields using Walsh sequences. These decoupling sequences act as digital filters that efficiently extract spectral coefficients while suppressing decoherence, thus providing improved sensitivity over existing strategies. 
We experimentally reconstruct the magnetic field radiated by a physical model of a neuron using a single electronic spin in diamond and discuss practical applications.
These results will be useful to implement time-resolved magnetic sensing with quantum probes at the nanometer scale.
\end{abstract}
\maketitle

\section{Introduction}
Measurements of weak electric and magnetic fields at the nanometer scale are indispensable in many areas, ranging from materials science to fundamental physics and biomedical science. In many applications, much of the information about the underlying phenomena is contained in the dynamics of the field.  While novel quantum probes promise to achieve the required combination of high sensitivity and spatial resolution, their application to efficiently mapping the temporal profile of the field is still a challenge. 

Quantum estimation techniques~\cite{Giovannetti11,Nielsen00b} can be used to measure time-varying fields by monitoring the shift in the resonance energy of a qubit sensor, e.g., via Ramsey interferometry. 
The qubit sensor, first prepared in an equal superposition of its eigenstates, accumulates a phase $\phi(T)=\gamma\int_0^Tb(t)dt$, where $\gamma$ is the strength of the interaction with the time-varying field $b(t)$ during the acquisition period $T$. The dynamics of the field could be mapped by measuring the quantum phase over successive, increasing acquisition periods~\cite{Balasubramanian09} or sequential small acquisition steps~\cite{Hall12}; 
however, these protocols are inefficient at sampling and reconstructing the field, as the former involves a deconvolution problem, while both are limited by short coherence times ($T_2^*$) that bound the measurement sensitivity.
Decoupling sequences~\cite{Carr54,Viola98,Khodjasteh05} could be used to increase the coherence time~\cite{Ryan10,Hall10b,Naydenov11,DeLange11}, but their application would result in a non-trivial encoding of the dynamics of the field onto the phase of the qubit sensor~\cite{Hirose12,Bylander11,Alvarez11,Bar-Gill12,Kotler11}.

\begin{figure}[h!t]
\centering
\includegraphics[width=0.45\textwidth]{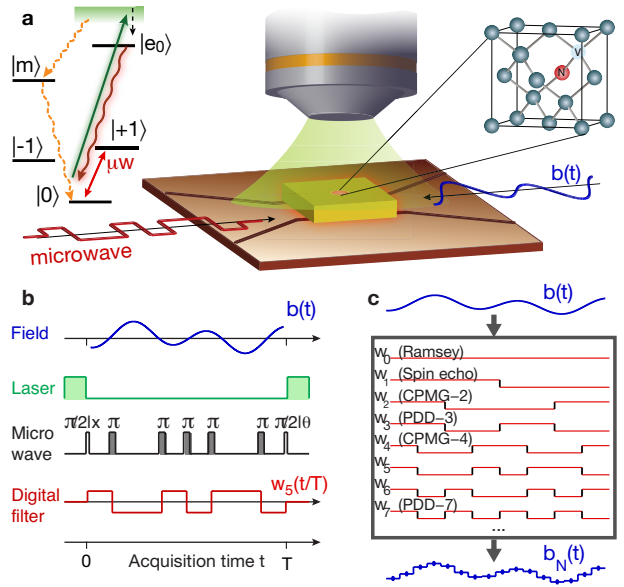}
\caption{\textbf{Walsh reconstruction protocol.} \textbf{a,} A single nitrogen-vacancy (NV) center in diamond, optically initialized and read out by confocal microscopy, is manipulated with coherent control sequences to measure the arbitrary profile of time-varying magnetic fields radiated by a coplanar waveguide under ambient conditions.  \textbf{b,}~Coherent control sequences, acting as digital filters on the evolution of the qubit sensor, extract information about time-varying fields.  \textbf{c,}~An $N$-point functional approximation of the field is obtained by sampling the field with a set of $N$ digital filters taken from the Walsh basis, which contain some known set of decoupling sequences such as the even-parity Carr-Purcell-Meiboom-Gill (CPMG) sequences~\cite{Carr54} ($w_{2^{n}}$) and the odd-parity Periodic Dynamical Decoupling (PDD) sequences~\cite{Khodjasteh05} ($w_{2^{n}-1}$).\label{fig1}}
\end{figure}

Instead, here we propose to reconstruct the temporal profile of time-varying fields by using
a set of digital filters, implemented with coherent control sequences over the whole acquisition period $T$, that simultaneously extract information about the dynamics of the field and protect against dephasing noise. 
In particular, we use control sequences~(Fig.~\ref{fig1}) associated with the Walsh functions~\cite{Walsh23}, which form a complete orthonormal basis of digital filters and are easily implementable experimentally. 

The Walsh reconstruction method can be applied to estimate various time-varying parameters via coherent control of any quantum probe.  {In particular, we show that the phase acquired by a qubit sensor modulated with Walsh decoupling sequences is proportional to the Walsh transform of the field}. 
This simplifies the problem of spectral sampling and reconstruction of time-varying fields by identifying the sequency domain as the natural description for dynamically modulated quantum systems.
{At the same time, the Walsh reconstruction method provides a solution to the problem of monitoring a time-dependent parameter with a quantum probe, which cannot be in general achieved via continuous tracking due to the destructive nature of quantum measurements.
In addition, because the Walsh reconstruction method achieves dynamical decoupling of the quantum probe, it further yields a significant improvement in coherence time and sensitivity over sequential acquisition techniques. 
These characteristics and the fact that the Walsh reconstruction method can be combined with data compression~\cite{Magesan13} and compressive sensing~\cite{Magesan13c, Candes06} provide clear advantages over prior reconstruction techniques~\cite{Balasubramanian09,Hall12,Schoenfeld11,Dreau11}.}


\section{Results}
\subsection{Walsh reconstruction method} 
The Walsh reconstruction method relies on the Walsh functions~(Supplementary Fig.~1), which are a family of piecewise-constant functions taking binary values, constructed from products of square waves, and forming a complete orthonormal basis of digital filters, analogous to the Fourier basis of sine and cosine functions.  The Walsh functions are usually described in a variety of labeling conventions, including the sequency ordering that counts the number of sign inversions or ``switchings'' of each Walsh function. The Walsh sequences are easily implemented experimentally by applying $\pi$-pulses at the switching times of the Walsh functions; these sequences are therefore decoupling sequences~\cite{Hayes11,Khodjasteh13}, which include the well-known Carr-Purcell-Meiboom-Gill (CPMG)~\cite{Carr54} and Periodic Dynamical Decoupling (PDD) sequences~\cite{Khodjasteh05}. 

Because a $\pi$-pulse effectively reverses the evolution of the qubit sensor, control sequences of $\pi$-pulses act as digital filters that sequentially switch the sign of the evolution between $\pm1$.  
If $w_m(t/T)$ is the digital filter created by applying $m$ control $\pi$-pulses at the zero-crossings of the $m$-th Walsh function, the normalized phase acquired by the qubit sensor is
\begin{equation}\label{eq1}
\frac{1}{\gamma T}\phi_m(T)=\frac{1}{T}\int_0^T b(t)w_m(t/T)dt\equiv\hat{b}(m).
\end{equation}
Here $\hat{b}(m)$ is the $m$-th Walsh coefficient defined as the Walsh transform of $b(t)$ evaluated at sequence number (sequency) $m$. 
This identifies the sequency domain as the natural description for digitally modulated quantum systems. Indeed, Eq.~(\ref{eq1}) implies a duality between the Walsh transform and the dynamical phase acquired by the qubit sensor under digital modulation, which allows for efficient sampling of time-varying fields in the sequency domain and direct reconstruction in the time domain via linear inversion.

Successive measurements with the first $N$ Walsh sequences $\{w_m(t/T)\}_{m=0}^{N-1}$ give a set of $N$ Walsh coefficients $\{\hat{b}(m)\}_{m=0}^{N-1}$ that can be used to reconstruct an $N$-point functional approximation to the field 
\begin{equation}\label{eq2}
b_{N}(t)=\sum_{m=0}^{N-1}\hat{b}(m)w_m(t/T).
\end{equation}
Eq.~(\ref{eq2}) is the inverse Walsh transform of order $N$, which gives the best least-squares Walsh approximation to $b(t)$. With few assumptions or prior knowledge about the dynamics of the field, the reconstruction can be shown to be accurate with quantifiable truncation errors and convergence criteria~\cite{Schipp90,Golubov91}.

Although the signal is encoded on the phase of a quantum probe in a different way (via decoupling sequences), the Walsh reconstruction method shares similarities with classical Hadamard encoding techniques in data compression, digital signal processing, and nuclear magnetic resonance imaging~\cite{Bolinger88,Kupce03,Tal10}. All of these techniques could easily be combined to achieve both spatial and temporal imaging of magnetic fields at the nanometer scale, given the availability of gradient fields and frequency-selective pulses.

\subsection{Walsh reconstruction of time-varying fields}
We experimentally demonstrate the Walsh reconstruction method by measuring increasingly complex time-varying magnetic fields.  We used a single NV center in an isotopically purified diamond sample as the qubit sensor~(details about the experimental setup can be found in Methods). 
{NV  centers in diamond~(Fig.~\ref{fig1}a) have recently emerged as promising sensors for magnetic~\cite{Taylor08, Maze08, Balasubramanian08} and electric~\cite{Dolde11} fields, rotations~\cite{Ledbetter12,Ajoy12g} and temperature~\cite{Toyli13, Kucsko13, Neumann13}. These sensors are ideal for nanoscale imaging of living biological systems~\cite{Chang08, McGuinness11, LeSage13} due to their low cytotoxicity, surface functionalizations~\cite{Mochalin12}, optical trapping capability~\cite{Horowitz12,Geiselmann13} and long coherence time under ambient conditions~\cite{Balasubramanian09}.}
A single NV center is optically initialized and read out by confocal microscopy under ambient conditions. 
A coplanar waveguide delivers both resonant microwave pulses and off-resonant time-varying magnetic fields produced by an arbitrary waveform generator.

\begin{figure}[t]
\centering
\includegraphics[width=0.4\textwidth]{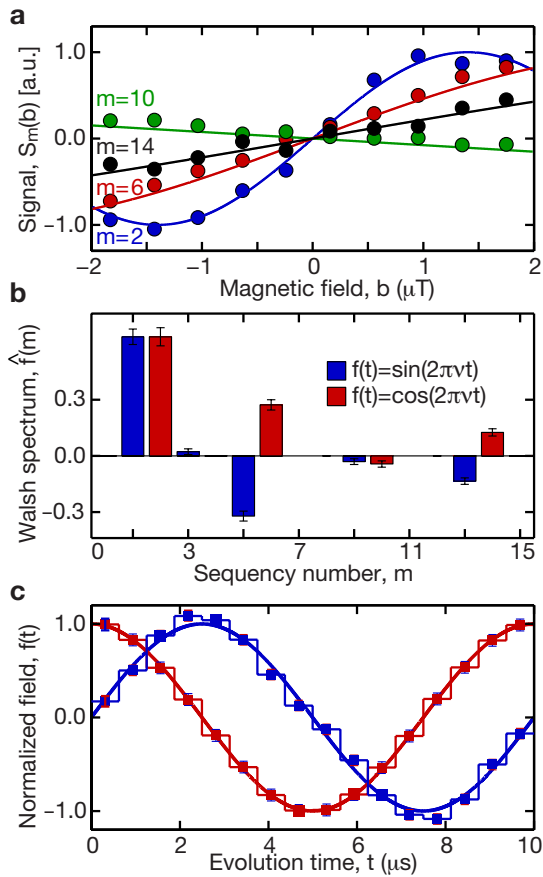}
\caption{\textbf{Walsh reconstruction of sinusoidal fields.} \textbf{a,}~Measured signal $S_m(b)=\sin(\gamma_eb\hat{f}(m)T)$ as a function of the amplitude of a cosine magnetic field for different Walsh sequences with $m$ $\pi$-pulses. Here $\gamma_e=2\pi\cdot28~\text{Hz}\cdot\text{nT}^{-1}$ is the gyromagnetic ratio of the NV electronic spin. The $m$-th Walsh coefficient $\hat{f}(m)$ is proportional to the slope of $S_m(b)$ at the origin. \textbf{b,}~Measured Walsh spectrum up to fourth order ($N=2^4$) of sine and cosine magnetic fields $b(t)\!=\!b\sin{(2\pi\nu t+\alpha)}$ with frequency $\nu=100$~kHz and phases $\alpha\in\{0,\pi/2\}$ over an acquisition period $T=1/\nu=10~\upmu\text{s}$. Error bars correspond to $95~\%$ confidence intervals on the Walsh coefficients associated with the fit of the measured signal. \textbf{c,}~The reconstructed fields (filled squares) are 16-point piecewise-constant approximations to the expected fields (solid lines, not a fit). Error bars correspond to the amplitude uncertainty of the reconstructed field obtained by propagation of the errors on the estimates of the uncorrelated Walsh coefficients.\label{fig2}}
\end{figure}

We first reconstructed monochromatic sinusoidal fields, $b(t)=b\sin{(2\pi\nu t+\alpha)}$, by measuring the Walsh spectrum up to fourth order ($N\!=\!2^4$). 
The $m$-th Walsh coefficient $\hat{f}(m)$ of the normalized field $f(t)=b(t)/b$ was obtained by sweeping the amplitude of the field and measuring the slope of the signal $S_m(b)=\sin{(\gamma b \hat{f}(m) T)}$ at the origin~(Fig.~\ref{fig2}a and Supplementary Fig.~2). Figure~\ref{fig2}b shows the measured non-zero Walsh coefficients of the Walsh spectrum. 
As shown in Fig.~\ref{fig2}c, the 16-point reconstructed fields are in good agreement with the expected fields.  We note that, contrary to other methods previously used for a.c. magnetometry, the Walsh reconstruction method is phase selective, as it discriminates between time-varying fields with the same frequency but different phase. 

\begin{figure}[t]
\centering
\includegraphics[width=0.4\textwidth]{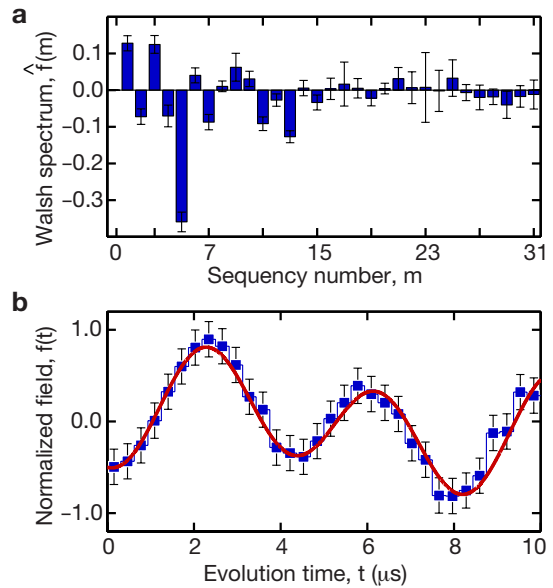}
\caption{\textbf{Walsh reconstruction of a bichromatic field.} \textbf{a,}~Measured Walsh spectrum up to fifth order ($N=2^5$) of a bichromatic magnetic field $b(t)\!=\!b\cdot[a_1\sin{(2\pi\nu_1t+\alpha_1)}+a_2\sin{(2\pi\nu_2t+\alpha_2)}]$ with $a_1=3/10$, $a_2=1/5$, $\nu_1=100$~kHz, $\nu_2=250$~kHz, $\alpha_1=-0.0741$, and $\alpha_2=-1.9686$. The zero-th Walsh coefficient $\hat{f}(0)$ corresponds to a static field offset that was neglected.  \textbf{b,}~The reconstructed field (filled squares) is a 32-point approximation to the expected field (solid line, not a fit). Error bars correspond to the amplitude uncertainty of the reconstructed field obtained by propagation of the errors on the estimates of the uncorrelated Walsh coefficients.\label{fig3}}
\end{figure}

We further reconstructed a bichromatic field $b(t)=b\,[a_1\sin{(2\pi\nu_1t+\alpha_1)}+a_2\sin{(2\pi\nu_2t+\alpha_2)}]$. Figure~\ref{fig3}a shows the measured Walsh spectrum up to fifth order ($N\!=\!2^5$). As shown in Fig.~\ref{fig3}b, the 32-point reconstructed field agrees with the expected field, which demonstrates the accuracy of the Walsh reconstruction method~(Supplementary Fig.~3). In contrast, sampling the field with an incomplete set of digital filters, such as the CPMG and PDD sequences, extracts only partial information about the dynamics of the field~(Supplementary Fig.~4).
By linearity of the Walsh transform, the Walsh reconstruction method applies to any polychromatic field (and by extension to any time-varying field), whose frequency spectrum lies in the acquisition bandwidth $[1/T,1/\tau]$  set by the coherence time $T\leq T_2$ and the maximum sampling time $\tau=T/N$, which is in turn limited by the finite duration of the control $\pi$-pulses.

\subsection{Performance of the Walsh reconstruction method}
The performance of the Walsh reconstruction method is determined by the reconstruction error $e_N$ and the measurement sensitivity $\eta_N$. The least-squares reconstruction error $e_N=\|b_N(t)-b(t)\|_2$ due to truncation of the Walsh spectrum up to $N\!=\!2^n$ coefficients is bounded by $e_N\leq\max_{t\in[0,T]}{|\partial_tb(t)|}/2^{n+1}$~\cite{Golubov91} and vanishes to zero as $N$ tends to infinity (as needed for perfect reconstruction). 
This implies that although the resources grow exponentially with $n$, the error converges exponentially quickly to zero, and only a finite number of coefficients is needed to accurately reconstruct the field. 

The measurement sensitivity of the $m$-th Walsh sequence in $M$ measurements,
\begin{equation}\label{eq3}
\eta_m=\frac{v_m^{-1}}{\gamma_eC\sqrt{T}|\hat{f}(m)|}=\frac{\hat{\eta}_m}{|\hat{f}(m)|},
\end{equation}
gives the minimum field amplitude, $\delta{b}_m=\eta_m/\sqrt{MT}=\Delta S_m/(|\partial S_m/\partial b_m|\sqrt{M})$, that can be measured with fixed resources. Here $\gamma_e=2\pi\cdot28~\text{Hz}\cdot\text{nT}^{-1}$ is the gyromagnetic ratio of the NV electronic spin and $C$ accounts for inefficient photon collection and finite contrast due to spin-state mixing during optical measurements~\cite{Taylor08,Aiello13}.

The sensitivity is further degraded by the decay of the signal visibility, $v_m=(e^{-T/T_2(m)})^{p(m)}\leq1$, where $T_2(m)$ and $p(m)$ characterize the decoherence of the qubit sensor during the $m$-th Walsh sequence in the presence of a specific noise environment. In general, $T_2(m)>T_2$, as the Walsh sequences suppress dephasing noise and extend coherence times by many orders of magnitude~\cite{Hayes11,Khodjasteh13}.
The sensitivity $\eta_m$ is thus the ratio between a field-independent factor $\hat{\eta}_m$ and the Walsh coefficient $|\hat{f}(m)|$ for the particular temporal profile of the measured field. 

{The sensitivity formula of Eq.~\ref{eq3} can be used to identify the Walsh sequences 
that extract the most information about the amplitude of time-varying fields in the presence of noise. In analogy to a.c. magnetometry~\cite{Taylor08}, which measures the amplitude of sinusoidal fields, we refer to the problem of performing parameter estimation of the amplitude of an arbitrary waveform as \textit{arbitrary waveform (a.w.) magnetometry}.
Indeed, if the dynamics of the field is known, the Walsh spectrum can be precomputed to identify the Walsh sequence that offers the best sensitivity. Because different Walsh sequences have different noise suppression performances~\cite{Hayes11,Magesan13}, the choice of the most sensitive Walsh sequence involves a trade-off between large Walsh coefficients and long coherence times.}

Measurements with an ensemble of $N_{\text{NV}}$ NV centers will improve the sensitivity by $1/\sqrt{N_{\text{NV}}}$, such that the sensitivity per unit volume will scale as $1/\sqrt{n_{\text{NV}}}$, where $n_{\text{NV}}$ is the density of NV centers. Previous studies have demonstrated $\eta_1\!\approx\!4~\text{nT}\cdot\text{Hz}^{-1/2}$ for a single NV center in an isotopically engineered diamond~\cite{Balasubramanian09} and $\eta_1\!\approx\!0.1~\text{nT}\cdot\text{Hz}^{-1/2}$ for an ensemble of NV centers~\cite{LeSage12}, with expected improvement down to $\eta_1\!\approx\!0.2~\text{nT}\cdot{\upmu\text{m}}^{3/2}\cdot\text{Hz}^{-1/2}$.

The amplitude resolution of the Walsh reconstruction method, $\delta b_N=\sqrt{\sum_m\delta \hat{b}_m^2}$, gives the smallest variation of the reconstructed field that can be measured from the Walsh spectrum of order $N$. If each Walsh coefficient is obtained from $M$ measurements over the acquisition period $T$, the measurement sensitivity of the Walsh reconstruction method, $\eta_N\equiv\delta b_N\sqrt{MNT}$, is 
\begin{equation}\label{eq4}
\eta_N=\sqrt{N\sum_m\hat{\eta}_m^2}=\frac{\sqrt{N\sum_{m=0}^{N-1}v_m^{-2}}}{\gamma_e C\sqrt{T}\sqrt{n_{\text{NV}}}}.
\end{equation}

The Walsh reconstruction method provides a gain in sensitivity of $\sqrt{N}$ over sequential measurement techniques that perform $N$ successive amplitude measurements over small time intervals of length $\tau=T/N\leq T_2^{*}$. Indeed, Walsh sequences exploit the long coherence time under dynamical decoupling to reduce the number of measurements, and thus the associated shot noise. {This corresponds to a decrease by a factor of $N$ of the total acquisition time needed to reach the same amplitude resolution or an improvement by a factor of $\sqrt{N}$ of the amplitude resolution at fixed total acquisition time~(see Supplementary Discussion).} Thus, unless the signal can only be triggered once, in which case one should use an ensemble of quantum probes to perform measurements in small time steps, the Walsh reconstruction method outperforms sequential measurements, which is an important step toward quantum-optimized waveform reconstruction~\cite{Tsang11}.

The measurement sensitivity $\eta_N$ combines with the reconstruction error $e_N$ to determine the accuracy of the Walsh reconstruction method.
If some small coefficients cannot be resolved due to low signal visibility, the increase in reconstruction error can be analytically quantified using data compression results~\cite{Magesan13}. In the same way, the acquisition time can be reduced by sampling only the most significant coefficients and discarding other negligible coefficients. Furthermore, if the field is sparse in some known basis, which is often the case, a logarithmic scaling in resources can be achieved by using compressed sensing methods based on convex optimization algorithms~\cite{Magesan13c,Candes06}, an advantage that is not shared by other sequential acquisition protocols~\cite{Balasubramanian09,Hall12,Schoenfeld11}.

\section{Discussion}
The Walsh reconstruction method is readily applicable to measure time-varying parameters in a variety of physical systems, including light shift spectroscopy with trapped ions~\cite{Kotler11}; magnetometry with single spins in semiconductors~\cite{Maze08,Koehl11,Pla13}
or quantum dots~\cite{DeGreve11}; and measurements of electric fields~\cite{Dolde11} or temperature~\cite{Toyli13, Kucsko13, Neumann13} with NV centers in diamond. Other promising applications include magnetic resonance spectroscopy of spins extrinsic to the diamond lattice~\cite{Staudacher13,Mamin13}, measurements of the dynamics of magnetic nanostructures~\cite{VanWaeyenberge06}, or magnetic vortices in nanodisk chains~\cite{Uhlir13}. 

{An active research direction for quantum sensors is measuring biological~\cite{Hall09,Hall10,LeSage13} and neuronal~\cite{Pham11, Hall12} activity at the nanometer scale}. 
Reference \cite{Hall12} provided compelling evidence about the feasibility of measuring the magnetic fields radiated by action potentials flowing through single neurons. They calculated magnetic field strengths of the order of $10~\text{nT}$ at distance up to $100~\text{nm}$ from a morphologically reconstructed hippocampal CA1 pyramidal neuron. These fields are within experimental reach using small ensembles of shallow-implanted NV centers, e.g., located less than $10~\text{nm}$ below the diamond surface~\cite{Ohno12, Staudacher12}, given improvements in collection efficiency~\cite{LeSage12,Marseglia11} and coherence times~\cite{Bar-Gill13}.

{The Walsh reconstruction method may also prove useful in neuroscience, alongside existing electrical activity recording techniques and other emerging neuroimaging modalities, to monitor the weak magnetic activity of neuronal cells at subcellular spatial resolution, as needed to better understand neurophysiology and map neuronal circuits. To achieve a repeatable signal and reduce stochastic fluctuations during averaging, sequential trains of action potentials could be evoked with conventional electrophysiological techniques, photo-stimulation methods, or current injection through underlying nanowire electrode arrays~\cite{Robinson12}. Technical issues associated with maintaing the stability of the system over long time scale still remain to be solved.

As a proof-of-principle implementation, we measured the magnetic field radiated by a physical model of a neuron undergoing an action potential $\Phi(t)$ approximated by a skew normal impulse~\cite{Wikswo80, Alle09, Debanne04}. Due to its linear response in the kHz regime~(Supplementary Fig.~5), our coplanar waveguide acts as the physical model of a neuron~(see Supplementary Methods), with the radiated magnetic field given by the derivative of the electric field~\cite{Swinney80,Woosley85}:~$b(t)=d\Phi(t)/dt$ (Supplementary Fig.~6).

\begin{figure}
\centering
\includegraphics[width=0.4\textwidth]{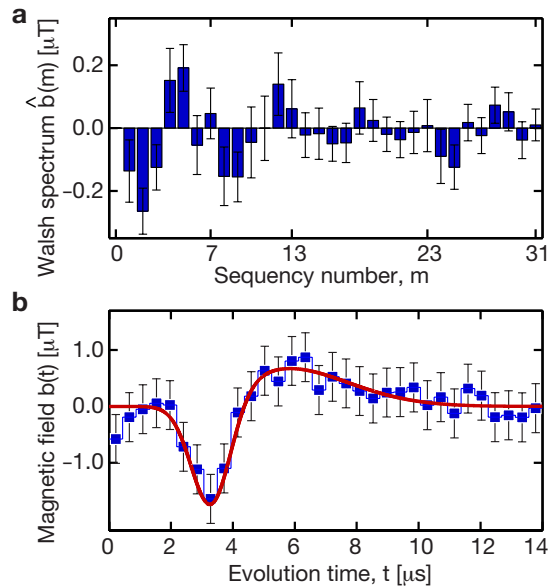}
\caption{\textbf{Walsh reconstruction of an arbitrary waveform.} \textbf{a,}~ Measured Walsh spectrum up to fifth order ($N=2^5$) of the magnetic field radiated by a skew normal impulse flowing through the physical model of a neuron. The Walsh coefficients were obtained by fixing the amplitude of the field and sweeping the phase of the last read-out $\pi/2$-pulse. {The acquisition time for measuring all the Walsh coefficients was less than $4$ hours.} Error bars correspond to $95~\%$ confidence intervals on the Walsh coefficients associated with the fit of the measured signal. \textbf{b,}~The reconstructed field (filled squares) is a 32-point approximation to the expected field (solid line, not a fit). Error bars correspond to the amplitude uncertainty of the reconstructed field obtained by propagation of the errors on the estimates of the uncorrelated Walsh coefficients. \label{fig4}}
\end{figure}

The Walsh coefficients were measured by fixing the amplitude of the field and sweeping the phase of the last read-out pulse to reconstruct the absolute field $b(t)$ rather than the normalized field $f(t)$. This protocol is in general applicable when the field amplitude is not under experimental control.
Figure~\ref{fig4}a shows the measured Walsh spectrum up to fifth order ($N=2^5$). As shown in Fig.~\ref{fig4}b, the $32$-point reconstructed field is in good agreement with the expected field. 
{Although neuronal fields are typically much smaller than in our proof-of-principle experiment with a single NV center, they could be measured with shallow-implanted single NVs~\cite{Ohno12, Staudacher12,Staudacher13,Mamin13} or small ensembles of NV centers~\cite{LeSage13,LeSage12,Pham11}.} 

In conclusion, we used control sequences acting as digital filters on the evolution of a single NV electronic spin to efficiently sample and accurately reconstruct the arbitrary profile of time-varying fields with quantifiable errors and formal convergence criteria. The Walsh reconstruction method can easily be used together with spatial encoding techniques to achieve both spatial and temporal imaging of magnetic fields. In addition, this method is compatible with data compression techniques~\cite{Magesan13} and compressed sensing algorithms~\cite{Magesan13c, Candes06} to achieve a significant reduction in resources, acquisition time, and reconstruction errors. Extension of the Walsh reconstruction method to stochastic fields could simplify the problem of spectral density estimation by removing the need for functional approximations or deconvolution algorithms~\cite{Biercuk09,Bylander11,Bar-Gill12}. This would enable, e.g., in-vivo monitoring of cellular functions associated with cell membrane ion-channel processes~\cite{Hall09,Hall10}. Finally, this work connects with other fields in which the Walsh functions have recently attracted attention, e.g., in quantum simulation to construct efficient circuits for diagonal unitaries~\cite{Welch13}, in quantum error suppression~\cite{Hayes11,Khodjasteh13}, and in quantum control theory to improve the fidelity of two-qubit entangling gates on trapped atomic ions~\cite{Hayes12}.

\section{Methods}
\subsection{Nitrogen-vacancy centers in diamond as qubit sensors}
Measurements of time-varying magnetic fields were performed under ambient conditions with a single NV center in an isotopically purified diamond sample ($>99.99\%$ C-12). 
The NV center in diamond consists of a substitutional nitrogen adjacent to a vacancy in the diamond lattice. The negatively charged defect exhibits a ground state electronic spin triplet. The zero-field energy splitting between the $m_s=0$ and $m_s=\pm1$ sublevels is $\Delta=2.87$~GHz. A static magnetic field $b_0=2.5$~mT directed along the quantization axis of the NV center lifts the energy degeneracy between the $m_s=+1$ and $m_s=-1$ sub-levels via the Zeeman effect. This gives an effective spin qubit $m_s=0\leftrightarrow m_s=1$ that can be used to measure time-varying magnetic fields. The strength of the interaction with external magnetic fields is given by the gyromagnetic ratio of the electron $\gamma_e=2\pi\cdot28~\text{Hz}\cdot\text{nT}^{-1}$. 
 The NV center was located in a home-built confocal microscope via fluorescence emission collection and optically initialized to its ground state with a $532$~nm laser pulse. 
 
Coherent control of the optical ground-state levels of the NV electronic spin was performed with resonant microwave pulses delivered through an on-chip coplanar waveguide. 
We set the effective Rabi frequency to $25$~MHz to achieve $20$~ns $\pi$-pulses and implemented phase cycling to correct for pulse errors.
The coplanar waveguide was also used to deliver non-resonant magnetic fields $\vec B(t)$ generated with an arbitrary waveform generator. The magnetometry method measured the resonance shift produced by the projection of the time-dependent field along the NV axis, $b(t)=\vec B(t)\cdot\vec{e}_z$. 

The qubit was optically read out via spin-state dependent fluorescence measurements in the $600-800$~nm spectral window around the $637$-nm zero-phonon line with a single-photon counter. 
 
\subsection{Walsh functions}
The set of Walsh functions~\cite{Walsh23,Schipp90,Golubov91} $\{w_m(t)\}_{m=0}^{\infty}$ is a complete, bounded, and orthonormal basis of digital functions defined on the unit interval $t\in[0,1[$. The Walsh basis can be thought of as the digital equivalent of the sine and cosine basis in Fourier analysis. 
The Walsh functions in the {dyadic ordering} or {Paley ordering} are defined as the product of Rademacher functions ($r_k$): $w_0=1$ and $w_m=\prod_{k=1}^{n}r_k^{m_k}$ for $1\leq m\leq2^n-1$, where $m_k$ is the $k$-th bit of $m$. The dyadic ordering is particularly useful in the context of data compression~\cite{Magesan13}.
The Rademacher functions are periodic square-wave functions that oscillate between $\pm1$ and exhibit $2^k$ intervals and $2^k-1$ jump discontinuities on the unit interval. Formally, the Rademacher function of order $k\geq1$ is defined as $r_k(t)\equiv r(2^{k-1} t)$, with
$$ r_k(t) = \left\{\begin{array}{ll} 1 & : t \in [0, 1/2^{k}[ \\ -1 & : t \in [1/2^{k}, 1/2^{k-1}[ \end{array} \right.$$  extended periodically to the unit interval. 
The Walsh functions in the {sequency ordering} are obtained from the gray code ordering of $m$. Sequency is a straightforward generalization of frequency which indicates the number of zero crossings of a given digital function during a fixed time interval. As such, the sequency $m$ indicates the number of control $\pi$-pulses to be applied at the zero crossings of the $m$-th Walsh function. The sequency ordering is thus the most intuitive ordering in the context of digital filtering with control sequences.
\\

\section{End notes}
\subsection{Acknowledgments}
We thank Jonathan Welch and Chinmay Belthangady for helpful discussions and Kurt Broderick for technical support. This work was supported in part by the U.S. Army Research Office through a MURI grant No. W911NF-11-1-0400 and by DARPA (QuASAR program). A. C. acknowledges support from the Natural Sciences and Engineering Research Council of Canada. 

\subsection{Author contributions statement}
A.C., E.M., and P.C. designed the method and did the theoretical analysis. A.C., H.N.Y., and P.C. designed and built the experimental system. A.C. and P.C. conducted the experiments, analysed the data and wrote the manuscript. All authors discussed the results and commented on the manuscript.

\subsection{Competing financial interests}
The authors declare no competing financial interests.

\newpage
\onecolumngrid


\section*{Supplementary Figures}
\vspace{-12pt}
\begin{figure}[h!] 
\centering
\includegraphics[scale=0.95]{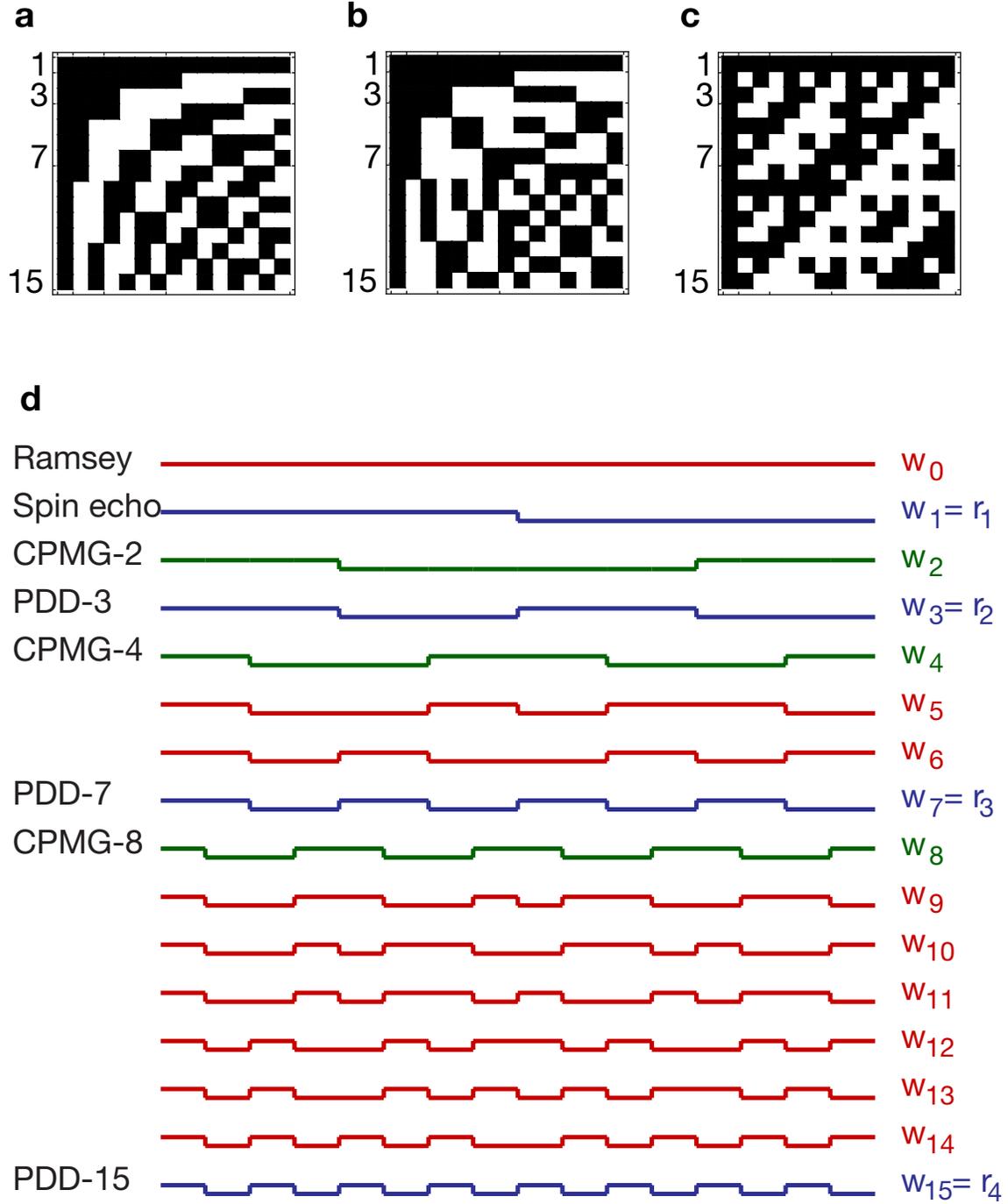}
\caption{\textbf{Walsh functions.}  Matrix representation of the Walsh functions up to fourth order ($N=2^4$) in \textbf{a,}~sequency ordering, \textbf{b,}~Paley ordering, and \textbf{c,}~Hadamard ordering. Each line corresponds to a Walsh sequence with the columns giving the value of the digital filter in the time domain. Black and white pixels represent the values $\pm1$. Each ordering can be obtained from the others by linear transformations.
\textbf{d} Walsh functions $\{w_m\}_{m=0}^{N-1}$ up to $N=2^4$ in sequency ordering. The sequency $m$ indicates the number of zero crossings of the $m$-th Walsh function. The Rademacher functions $r_k=w_{2^{k}-1}$ correspond to the Walsh functions plotted in blue. Some Walsh functions are associated with known decoupling sequences such as the even-parity CPMG sequences~\cite{Carr54} ($w_{2^{k}}$ green lines) and the odd-parity PDD sequences~\cite{Khodjasteh05} ($w_{2^{k}-1}$, blue lines).}
\label{som_walsh_fig3}
\end{figure}

\newpage
\begin{figure}[h!] 
\centering
\includegraphics[scale=0.95]{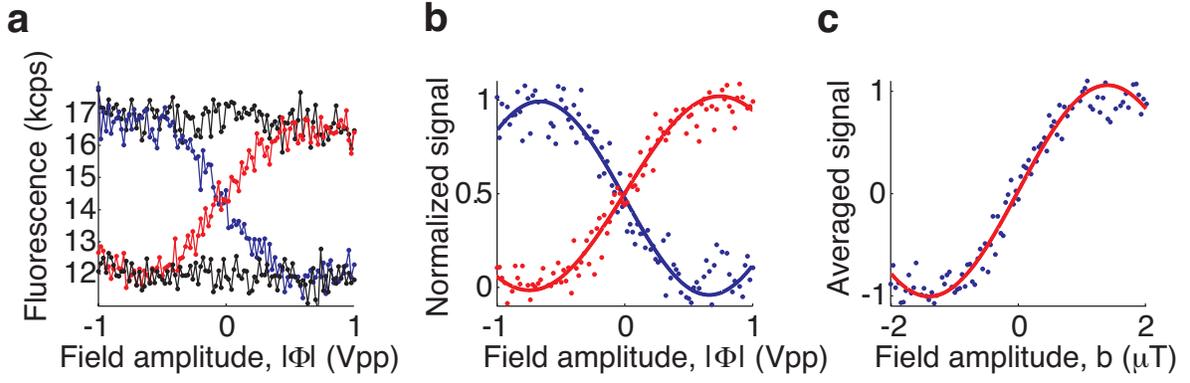}
\caption{\textbf{Raw experimental data.} Example of experimental data for measuring the $m$-th Walsh coefficient. \textbf{a,}~Measured fluorescence signals $S_{xy}$ and $S_{x\bar{y}}$ for a sinusoidal field measured with the $m=1$ Walsh sequence. \textbf{b,}~The measured fluorescence signals are normalized with respect to the reference signals. \textbf{c,}~Average normalized fluorescence signal $S_m(b)$ whose slope at the origin is proportional to the $m$-th Walsh coefficient $\hat{f}(m)$ of the normalized field $f(t)$.}
\label{som_calibration_fig2}
\end{figure}

\newpage
\begin{figure}[h!] 
\centering
\includegraphics[scale=0.95]{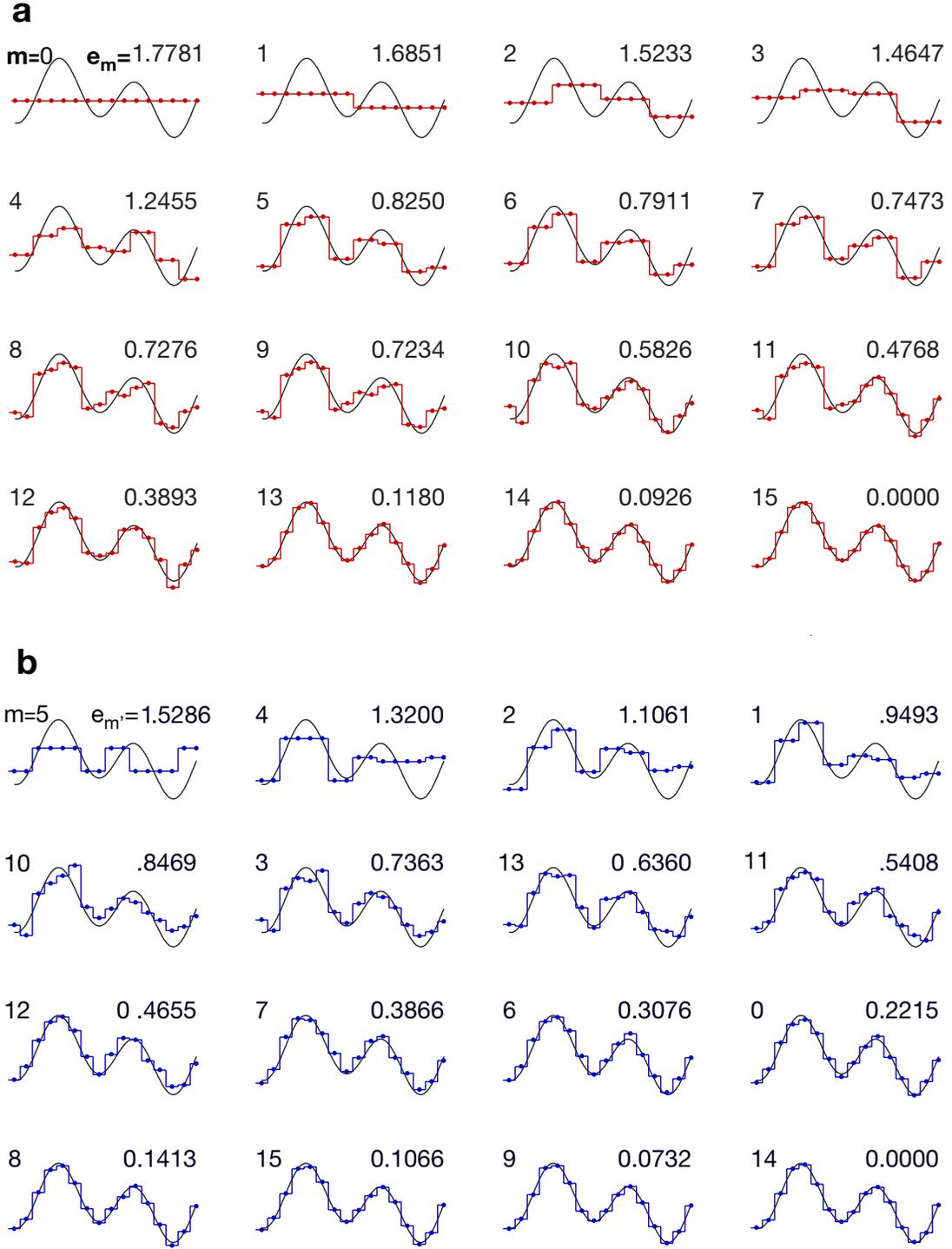}
\caption{\textbf{Accuracy of the Walsh reconstruction method.} \textbf{a}. Reconstruction of a bichromatic field with an increasing number of Walsh coefficients $\{\hat{f}(m)\}_{m=0}^{m_{max}}$ in {sequency} ordering.   \textbf{b}. Reconstruction of a bichromatic field with an increasing number of Walsh coefficients $\{\hat{f}(m)\}_{m=0}^{m'_{max}}$  sorted by the coefficient magnitude. A finite number of coefficients is needed to reconstruct an accurate estimate of the field. The upper left and upper right numbers respectively correspond to $m$ (in the sequence order) and the $l_2$-reconstruction error up to the $m'$ reconstruction, $e_{m'}$.}
\label{som_reconstruction_fig2}
\end{figure}

\newpage
\begin{figure}[h!] 
\centering
\includegraphics[width=0.8\textwidth]{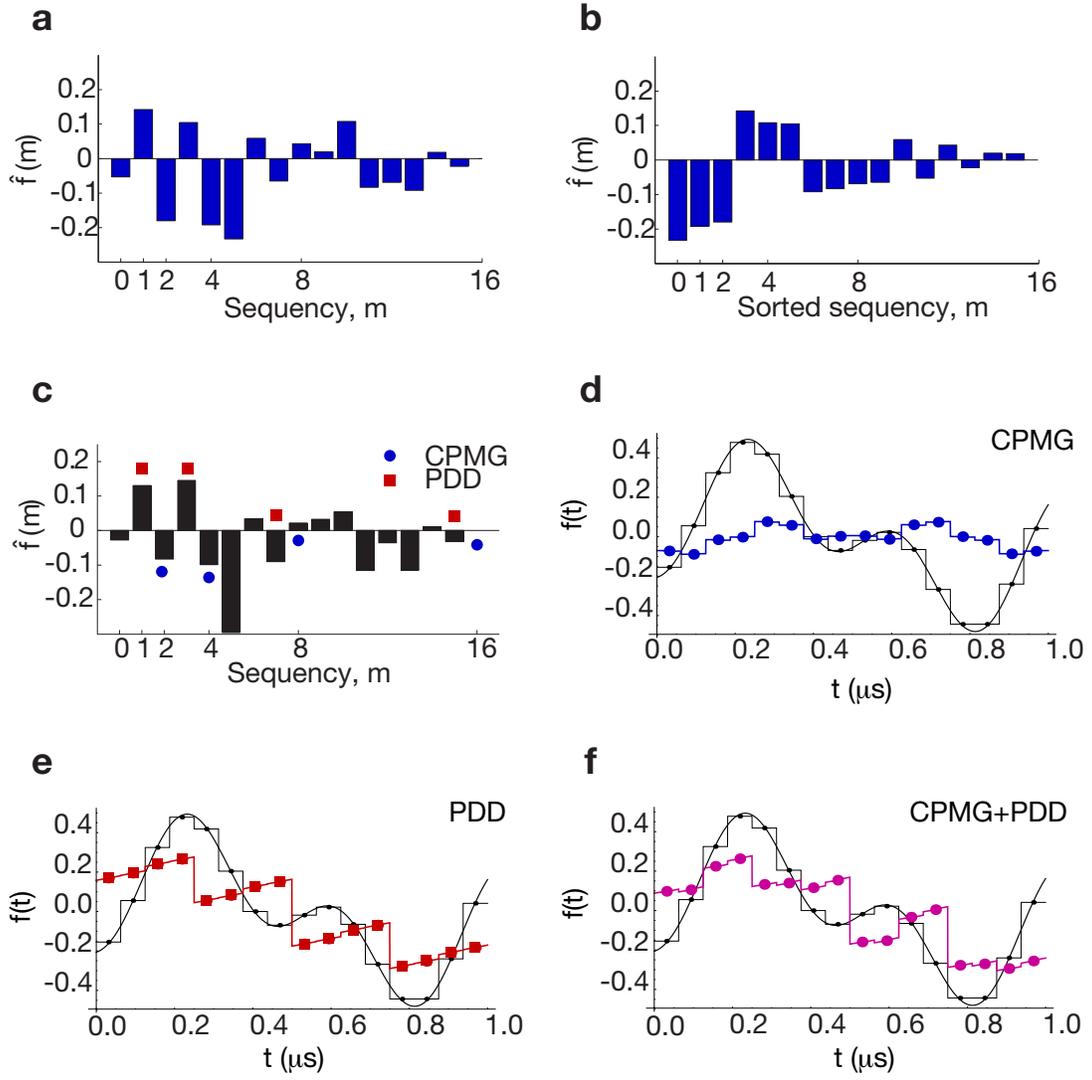}
\caption{\textbf{Reconstruction of a bichromatic field with Walsh functions.}  Simulated Walsh spectrum for a bichromatic field. Walsh coefficients are presented in \textbf{a,}~sequency ordering and  \textbf{b,}~sorted ordering with decreasing magnitude. If the temporal profile of the field is known, the resources can be allocated to extract the most significant information about the field by sampling the largest Walsh coefficients.  \textbf{c, } Walsh spectrum up to fifth order $(N=2^4=16)$. The blue dots and red squares respectively indicate the subset of coefficients associated with the PDD sequences ($w_{2^{n}-1}$) and CPMG sequences ($w_{2^{n}}$). The field reconstructed with the first 16 CPMG coefficients (\textbf{d,}),   16 PDD coefficients (\textbf{e,}) and both PDD and CPMG coefficients (\textbf{f,}),  remains inaccurate in comparison with the reconstructed field obtained with the first 16 Walsh coefficients (black solid line in \textbf{d-f}).}
\label{som_reconstruction_fig3}
\end{figure}

\newpage
\begin{figure}[h!]
\centering
\includegraphics[scale=0.95]{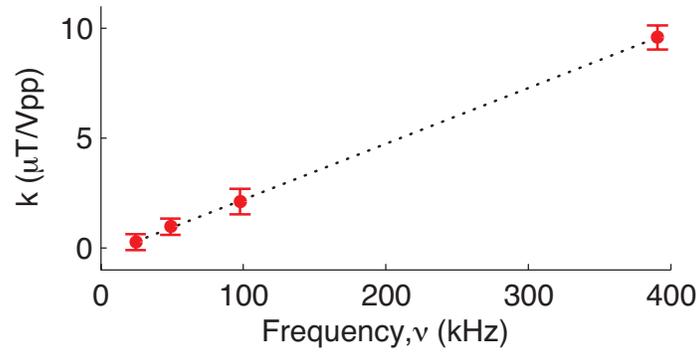}
\caption{\textbf{Calibration of the coplanar waveguide.} The conversion factor converts the amplitude of the electric field $\Phi(t)$ into the amplitude of the magnetic field $b(t)$ measured by the NV center. The conversion factor depends linearly on the frequency. Error bars are standard deviation of measurements.}
\label{som_calibration_fig1}
\end{figure}

\newpage
\begin{figure}[h!] 
\centering
\includegraphics[scale=0.95]{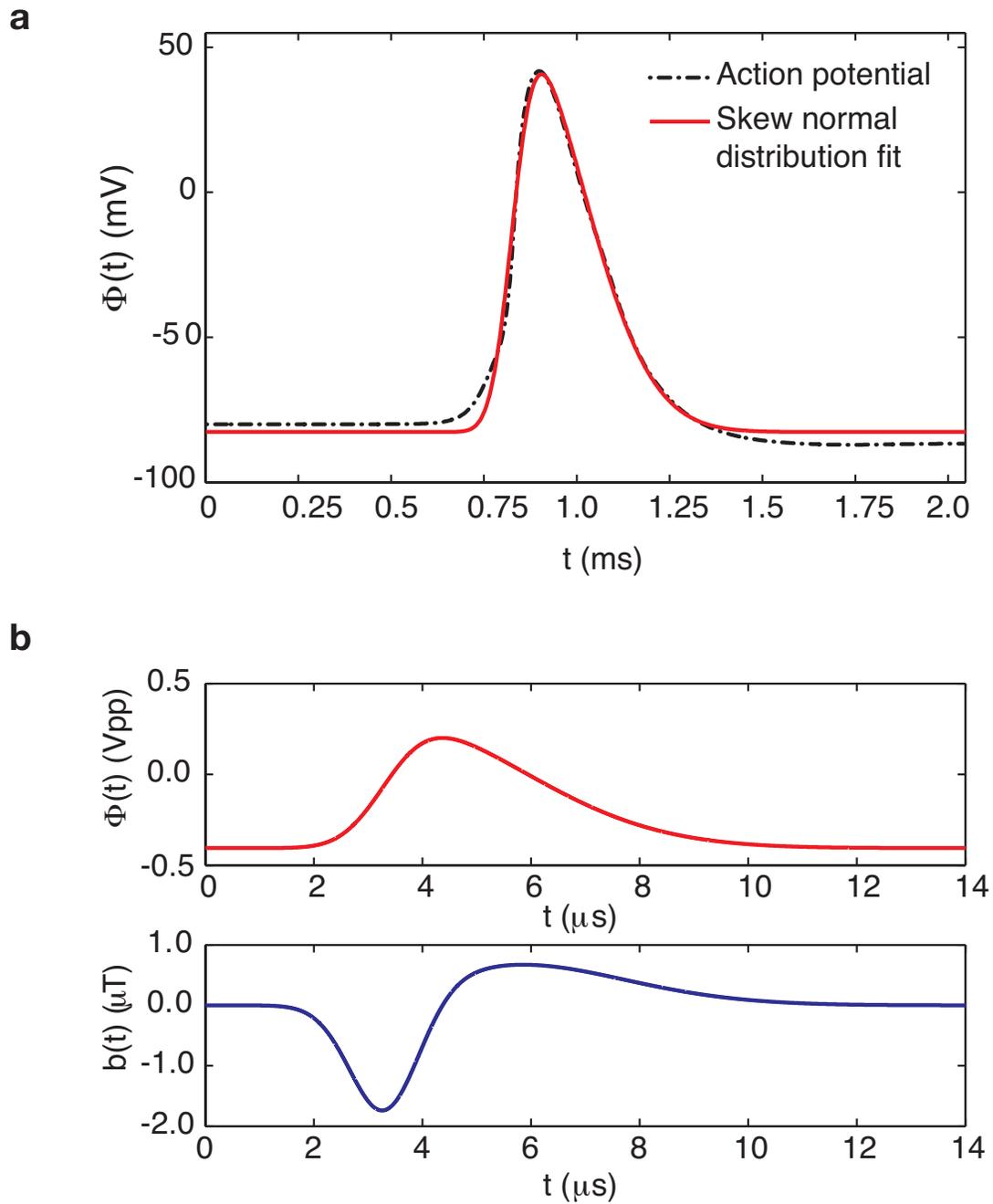}
\caption{\textbf{Simulated action potential.} \textbf{a} ,The simulated action potential of a rat hippocampal mossy fiber boutons~\cite{Alle09} is approximated by a skew normal impulse. \textbf{b}, The action potential is rescaled to perform the Walsh reconstruction experiment, with the radiated magnetic field corresponding to the first derivative of the action potential.}
\label{som_neuronal_fig1}
\end{figure}

\newpage

\section*{Supplementary Discussion}
\subsection{Walsh functions}
The set of Walsh functions~\cite{Walsh23,Schipp90,Golubov91} $\{w_m(t)\}_{m=0}^{\infty}$ is a complete, bounded, and orthonormal basis of digital functions defined on the unit interval $t\in[0,1[$. The Walsh basis can be thought of as the digital equivalent of the sine and cosine basis in Fourier analysis.  There are different orderings of the Walsh functions in the basis that are interchangeably used depending on the various conventions adopted in different fields.
The matrix representations of the Walsh functions in three orderings are compared in Supplementary Fig.~\ref{som_walsh_fig3}.

The Walsh functions in the \emph{dyadic ordering} or \emph{Paley ordering} are defined as the product of Rademacher functions~($r_k$): $w_0=1$ and $w_m=\prod_{k=1}^{n}r_k^{m_k}$ for $1\leq m\leq2^n-1$, where $m_k$ is the $k$-th bit of $m$. The dyadic ordering is particularly useful in the context of data compression~\cite{Magesan13}.

The Rademacher functions are periodic square-wave functions that oscillate between $\pm1$ and exhibit $2^k$ intervals and $2^k-1$ jump discontinuities on the unit interval. Formally, the Rademacher function of order $k\geq1$ is defined as $r_k(t)\defeq r(2^{k-1} t)$, with
$$ r(t) = \left\{\begin{array}{ll} 1 & : t \in [0, 1/2[ \\ -1 & : t \in [1/2, 1[ \end{array}   \right.~\text{ and}~ r_k(t) = \left\{\begin{array}{ll} 1 & : t \in [0, 1/2^{k}[ \\ -1 & : t \in [1/2^{k}, 1/2^{k-1}[ \end{array} \right.$$  extended periodically to the unit interval. 

The Walsh functions in the \emph{sequency ordering}~(Supplementary Fig.~\ref{som_walsh_fig3}.d) are obtained from the gray code of $m$. Sequency is a straightforward generalization of frequency which indicates the number of zero crossings of a given digital function during a fixed time interval. As such, the sequency $m$ indicates the number of control $\pi$-pulses to be applied at the zero crossings of the $m$-th Walsh function. The sequency ordering is thus the most intuitive ordering in the context of digital filtering with control sequences. 

The Walsh functions in the \emph{Hadamard ordering} are represented by the Walsh-Hadamard square matrix of size $2^n \times 2^n$, whose elements are given by $H^{(n)}(i+1,j+1)=\prod_{l=0}^{2^n-1}(-1)^{i_l \cdot j_l}$. The Walsh-Hadamard matrix is used as a quantum gate in quantum information processing to prepare an equal superposition of $2^n$ orthogonal states from a set of $n$ initialized qubits. 

\subsection{Walsh transform}
The \emph{Walsh-Fourier series} of an integrable function $b(t)\in L^1([0,T[)$ is given by
\begin{equation}
b(t)\defeq\sum_{m=0}^\infty\hat{b}(m)w_{m}(t/T),
\end{equation}
where the \emph{Walsh-Fourier coefficients} are given by the Walsh transform of $b(t)$ evaluated at sequency $m$, i.e., 
\begin{equation}
\hat{b}(m)=\frac{1}{T}\int_0^Tb(t)w_m(t/T)dt\in\mathbb{R},~\forall m\geq0.
\end{equation}

The truncation of the Walsh-Fourier series up to $N$ coefficients gives the $N$-th partial sums of the Walsh-Fourier series,
\begin{equation}\label{eq3}
b_N(t)\defeq\sum_{m=0}^{N-1}\hat{b}(m)w_{m}(t/T),
\end{equation}
which can be shown to satisfy $\lim_{N\rightarrow\infty}b_N(t)=b(t)$, almost everywhere for $b(t)\in L^1([0,T[)$ and uniformly for $b(t)\in C([0,T[)$~\cite{Golubov91}. Supplementary equation~(\ref{eq3}) can be associated with an $N$-point functional approximation to the field $b(t)$. 

\subsection{Sensitivity of the Walsh reconstruction method}
While in the main text we focused on applying Walsh sequences to reconstruct the temporal profile of time-varying magnetic fields, we note that for time-varying fields with known dynamics, Walsh sequences provide a systematic way to choose for the control sequence that gives the best estimate of the amplitude of the field with the optimal sensitivity.  For example, the spin-echo sequence ($w_1(t/T)$) gives the best sensitivity for measuring the amplitude of a sinusoidal field $b(t)=b\sin{(2\pi\nu t)}$ with known frequency $\nu=1/T$. If the dynamics of the field is only partially known, an estimate of the field amplitude can be obtained by distributing the resources over a fixed subset of Walsh sequences; for example, by choosing the spin-echo sequence ($w_1(t/T)$) and CPMG-2 sequence ($w_2(t/T)$) to measure the amplitude of an oscillating field $b(t)=b\sin{(2\pi\nu t+\alpha)}$ with known frequency $\nu=1/T$ but unknown phase $\alpha$.

We can compute the minimum field amplitude that can be estimated from $M$ measurements with the $m$-th Walsh sequence from the quantum Cramer-Rao bound:
$$\delta b_m=\frac{1}{\sqrt{M}}\frac{\Delta S_m}{|\partial S_m/\partial b|}=\frac{v_m^{-1}}{\gamma_e C\sqrt{M}T|\hat{f}(m)|},$$
where $\gamma_e=2\pi\cdot28~\text{Hz}\cdot\text{nT}^{-1}$ is the gyromagnetic ratio of the NV electronic spin,  $C$  accounts for inefficient photon collection and finite contrast due to spin-state mixing during optical measurements~\cite{Taylor08,Aiello13} and $v_m=(e^{-T/T_2(m)})^{p(m)}\leq1$ is the visibility,  which depends on the parameters $T_2(m)$ and $p(m)$, which characterize the decoherence of the qubit sensors during the $m$-th Walsh sequence.

The minimum field amplitude $\delta b_m$ is related to the minimum resolvable Walsh coefficient $\delta \hat b_m=\Delta S_m/|\partial S_m/\partial \hat b_m|\sqrt{M}=\delta b_m|\hat{f}(m)|$. Given a total measurement time $T_{total}=MT$, the corresponding sensitivities are $\eta_m=\delta b_m\sqrt{\mathcal{T}}$ and $\hat\eta_m=\delta\hat b_m\sqrt{\mathcal{T}}=\eta_m|\hat{f}(m)|$~(cf. Eq.~(3) of the main text).
The statistical error on the field function $b_N(t)$ reconstructed from  the measured set of $N$ Walsh coefficients $\{\hat{b}(m)\}_{m=0}^{N-1}$ is obtained from the errors on each coefficient by $\delta b_N^2(t)=\sum_m \delta\hat b_m^2 |w_m(t)|^2=\sum_m \delta\hat b_m^2\mathbbm{1}_{[0,T[}(t)$, which is constant on the time interval $t\in[0,T[$. This quantity gives the measurement sensitivity of the Walsh reconstruction method $\eta_N=\delta b_N\sqrt{N\mathcal{T}}=\sqrt{N\sum_m\hat{\eta}_m^2}$~(cf. Eq. (4) of the main text).

\subsection{Sensitivity improvement over existing methods}\label{sec:etacompare}
We consider the problem of reconstructing an $N$-point approximation to the time-varying field $b(t)$ over the acquisition period of length $T$. Here we show that the Walsh reconstruction method offers an improvement in sensitivity scaling as $\sqrt{N}$ over existing sequential methods. This corresponds to a reduction by $\sqrt{N}$ of the minimum detectable field at fixed total acquisition time or a reduction by $N$ of the total acquisition time at fixed minimum detection field.
We compare the Walsh reconstruction method to piece-wise reconstruction of the field via successive acquisition (e.g. with a Ramsey~\cite{Balasubramanian09,Hall12} or CW~\cite{Schoenfeld11} method). 

Consider first that we use sequential (e.g., Ramsey) measurements to estimate the amplitude of the field during each of the $N$ sub-intervals of length $\tau=T/N$. The error on each point for shot-noise limited measurements scales as $\delta b_j\sim1/\tau\sqrt{M_1}$, where $M_1$ is the number of measurements performed for signal averaging. The total acquisition time is $\mathcal{T}_1=M_1 N \tau=M_1 T$ and the total reconstruction error on the $N$-point reconstructed field is 
\begin{equation}
(\delta b_N)_1=\sqrt{\sum_{j=0}^{N-1}\delta b_j^2}=\sqrt{N}\delta b_j\sim\sqrt{N}/\tau\sqrt{M_1}.
\end{equation} 

We compare this result to Walsh reconstruction performed via $N$ Walsh measurements, each over the acquisition period $T$. The error on each Walsh coefficient for shot-noise limited measurements scales as $\delta \hat{b}_m\sim1/T\sqrt{M_2}$, where $M_2$ is the number of measurements performed for signal averaging. The total acquisition time is $\mathcal{T}_2=M_2 N T$ and the total reconstruction error on the $N$-point reconstructed field (	 the main text) is 
\begin{equation}
(\delta b_N)_2=\sqrt{\sum_{m=0}^{N-1}\delta \hat{b}_m^2}=\sqrt{N}\delta \hat{b}_m\sim\sqrt{N}/T\sqrt{M_2}.
\end{equation}

For a fixed total acquisition time, $\mathcal{T}_1=M_1 T=M_2 N T=\mathcal{T}_2$, we have $M_1=N M_2$, which means that $N$ times more Ramsey measurements can be performed than Walsh measurements. We however have $(\delta b_N)_1=\sqrt{N}(\delta b_N)_2$, which means that the Walsh measurements is more sensitive by a factor of $\sqrt{N}$.

For a fixed minimum detectable field, $(\delta b_N)_1\sim\sqrt{N}/\tau\sqrt{M_1}=\sqrt{N}/T\sqrt{M_2}\sim(\delta b_N)_2$, we have $M_1=N^2 M_2$, which means that $N^2$ times more Ramsey measurements are needed to measure the exact same minimum field provided by $N$ Walsh measurements. We have then $\mathcal{T}_1=N\mathcal{T}_2$, such that the total acquisition time for Ramsey measurements is $N$ times longer than for Walsh measurements.

In summary, the Walsh reconstruction method offers a $\sqrt{N}$ improvement in sensitivity over Ramsey measurements, i.e. $(\eta_N)_2=(\eta_N)_1/\sqrt{N}$.

Further improvement in sensitivity can be achieved with data compression~\cite{Magesan13} and compressive sensing~\cite{Magesan13c} techniques. Given $N_2<N_1$ and $T=N_1\tau$, we find
\begin{equation}
\frac{(\eta_N)_1}{(\eta_N)_2}=\frac{N_1}{N_2}\sqrt{\frac{T}{\tau}}=\frac{N_1}{N_2}\sqrt{N_1}
\end{equation}

For $N_2=N_1=N$, we retrieve $(\eta_N)_2=(\eta_N)_1/\sqrt{N}$. Given a compression rate $\kappa<1$, such that $N_2=\kappa N_1$, we find $(\eta_N)_2=\kappa (\eta_N)_1/\sqrt{N_1}<(\eta_N)_1/\sqrt{N_1}$. For compressive sensing with resources scaling as $N_2\sim\log_2(N_1)$, we find

\begin{equation}
(\eta_N)_2\sim\frac{\log_2(N_1)}{N_1}\frac{(\eta_N)_1}{\sqrt{N_1}}<\frac{(\eta_N)_1}{\sqrt{N_1}}.
\end{equation}

In the discussion above, we did not consider other error sources besides shot noise. Indeed, while the sequential acquisition times are limited by the dephasing noise, the coherence time under the Walsh sequence is in general much longer (typically by 2-3 orders of magnitude for NV centers), thus additional decoherence losses are comparable in the two protocols. 
We note that even if the total acquisition time $T$ over which one wants to acquire the signal were longer than the coherence time under Walsh decoupling, $T_2$, it would still be advantageous to use sequential Walsh reconstruction over smaller time intervals,~$\sim T_2$. Other considerations, such as power broadening in CW experiments and dead-times associated with measurement and initialization of the quantum probe, make the Walsh reconstruction method even more advantageous.

\subsection{Reconstruction accuracy}
The Walsh spectrum of order $N=16$ for a bichromatic field is shown in Supplementary Fig.~\ref{som_reconstruction_fig3}a-b. The Walsh coefficients were first computed in term of their sequency $m$, which is related to the number of $\pi$ pulses needed to implement the corresponding $m$-th Walsh sequence. The amount of information provided by each Walsh sequence is proportional to the magnitude of the Walsh coefficient; the Walsh spectrum can thus be reordered with the coefficients sorted in decreasing order of their magnitude (Supplementary Fig.~\ref{som_reconstruction_fig3}b). As each Walsh coefficient adds new information to the reconstructed field, the accuracy of the reconstruction improves while the reconstruction error decreases; this is shown in Supplementary Fig.~\ref{som_reconstruction_fig2}.

Supplementary Fig.~\ref{som_reconstruction_fig2}a shows the reconstructed field with an increasing number of Walsh coefficients ($m\in[0,m_{max}]$) in sequency ordering.  The reconstruction error decreases monotonically as the number of coefficients increases. If the dynamics of the field is known, the Walsh coefficients can be precomputed and sorted so to allocate the available resources to sample the largest coefficients, which provide the most information about the field.  This is shown in Supplementary Fig.~\ref{som_reconstruction_fig2}b, where not only the accuracy of the reconstruction improves monotonically, but also only the first few coefficients are needed to achieve an accurate estimate of the field. 

\subsection{Reconstruction accuracy improvement over finite sampling with digital decoupling sequences}
The set of Walsh sequences contains some known set of digital decoupling sequences such as the Carr-Purcell-Meiboom-Gill (CPMG) sequences ($w_{2^{n}}$, $n\geq1$) and the periodic dynamical decoupling (PDD) sequences ($w_{2^{n}-1}$, $n\geq1$), which have been studied in the context of dynamical error suppression and noise spectrum reconstruction. 

Although the CPMG and PDD sequences can be shown to contain some significant information about the field, they do not contain all the significant information. The CPMG and PDD sequences are indeed symmetric and anti-symmetric functions about their midpoint; they only sample the even and odd symmetries of the field. This is shown in Supplementary Fig.~\ref{som_reconstruction_fig3}c-f where the Walsh reconstruction method outperforms the CPMG and PDD sequences, even if the same amount of resources is allocated to sample an equal number of  coefficients; because the field does not have a definite parity, the CPMG and PDD sequences fail to accurately reconstruct the field.  In addition, these sequences require an exponentially large number of control pulses, which may be detrimental in the presence of pulse errors. Therefore, sampling the field with only these sequences provides incomplete information about the Walsh spectrum and thus leads to inaccurate reconstruction in the time domain. 

CPMG and PDD have been used as filters in the frequency domain to achieve frequency-selective detection and reconstruction of noise spectral density. Even for this task, Walsh reconstruction can provide an advantage. Indeed, in the frequency domain, digital filters are trigonometric functions that are not perfectly approximated by delta functions and exhibit spectral leakage, i.e., the non-zero side-lobes of the filter function capture non-negligible signal contributions about other frequencies than the main lobe. Although the CPMG and PDD sequences can be tuned to sample the field at a specific central frequency, they also capture signal at other frequencies, which prevents the accurate reconstruction of time-varying fields. The Walsh reconstruction method removes the need for functional approximations or deconvolution algorithms by choosing the representation that is natural for digital filters: the Walsh basis.

\subsection{Comparison with prior reconstruction methods}
The problem of measuring the time-varying magnetic fields with NV centers in diamond has also been discussed by~ Balasubramanian \textit{et al.}~\cite{Balasubramanian09} and Hall \textit{et al.}~\cite{Hall12}. They both considered Ramsey interferometry measurements to detect time-varying magnetic fields, either by sweeping the measurement time or translating the acquisition window.

The first protocol presents a deconvolution problem that requires numerical algorithms to be solved and makes it difficult to analytically quantify the reconstruction error. In addition, the field is not efficiently sampled due to spectral leakage, because the window function is a sinc function, rather than a delta function, in the frequency domain. The Walsh transform method simplifies the problem of spectral sampling and reconstruction by setting the sequency domain, rather than the frequency domain, as the natural description for digitally sampled fields.

The second protocol does not require solving the inversion problem, as it directly gives the average field value during each sampling interval. However, the time required to achieve the same sensitivity as the Walsh method for an $N$-point reconstruction is $N$ times longer.

More importantly, these methods are not compatible with optimized protocols based on adaptive and compressive sampling. In particular, compressive sensing is fully compatible with the Walsh reconstruction method and  for many sparse signals it would lead to a large saving in measurement time. Indeed, sequentially measuring the field over small time intervals $\delta t$ requires performing $N=T/\delta t$ acquisitions to reconstruct the whole evolution, while one in general only needs $m\propto\log(N)$ measurements using compressive sensing techniques~\cite{Magesan13c}.

Hall \textit{et al.}~\cite{Hall12} also considered using an optically detected magnetic resonance protocol~\cite{Schoenfeld11,Dreau11} for measuring time-varying fields. This protocol is meant to perform real-time measurements of the resonance frequency of the NV centers via a lock-in detection system. Although this protocol is presented as continuously monitoring the field, a finite binning time is required ($\sim24~\upmu\text{s}$ in the case of Ref.~\cite{Hall12}) since the measurement on the NV center is destructive. In addition, continuous driving protocols are inherently less sensitive than pulsed ones, as the continuous laser light and microwave field induce power broadening, in addition to heating, which can be detrimental to biological samples.

\section*{Supplementary Methods}
\subsection{Measurement setup}
We conduct experiments on a single nitrogen-vacancy (NV) center in an isotopically purified diamond layered sample (Element Six, $>99.99\%$ C-12). The diamond sample is mounted on a three-axis nanopositioning stage (MCL NANO-LP200) and single NV centers are localized by fluorescence measurements in a home-built confocal microscope. A continuous $532$~nm laser light (Lighthouse Photonics Sprout-5) modulated with an acoustic-optic modulator (IntraAction AFM-8041) optically excites the NV center. The fluoresence light emitted is collected by an oil-immersion objective (NA=1.3, Nikon MRH01902) and filtered through a dichroic mirror (Chroma ZT532RDC) and long-pass edge filter (Semrock BLP01-594R-25). The photons are detected with a photon counter (Perkin Elmer SPCM-AQRH-15-FC) connected to a data acquisition card (National Instruments PCIe-6323). 

We perform optically detected magnetic resonance experiments under ambient conditions. A signal generator (SRS SG384) generates an oscillating radiofrequency current that is phase modulated with an IQ mixer (Marki microwave IQ-0318), gated with a microwave switch (Mini-Circuits ZASWA-2-50DR), and amplified (Mini-Circuits ZHL-16W-43+). The electric pulses are sent through an on-chip copper coplanar waveguide, which radiates a magnetic field at the location of the NV center. An arbitrary waveform generator (Tektronix AWG5014C) produces digital pulses that control and trigger the various electronic devices. Two analog channels generate continuous sinusoidal fields oscillating in phase quadrature  at 250MHz to modulate the IQ mixer. A third analog channel generates electric fields in the kHz frequency range that are amplified (Mini-Circuits, ZHL-32A BNC) before being sent through the other port of the coplanar waveguide, with a low-pass filter (Mini-Circuits VLF-180 SMA) and high-pass filter (Mini-Circuits VHF-1300) on both side of the coplanar waveguide to protect the amplifiers. The decrease in amplification gain at frequencies lower than $50$~kHz prevents the generation of fields over long acquisition periods.  

\subsection{Calibration of the coplanar waveguide}

Coplanar waveguides were fabricated by e-beam photolithography on microscope glass coverslips, soldered on a PCB board, and mounted to the confocal microscope. Electric waveforms $\Phi(t)$~(Vpp) were generated with an arbitrary waveform generator, amplified, and sent through the coplanar waveguide. The coplanar waveguide radiates a magnetic field $b(t)$~(nT) at the location of the NV center which can be derived from $\Phi(t)$ via a conversion factor $k(\nu)~(\text{nT}\cdot\text{Vpp}^{-1})$.

We calibrated the conversion factor by performing a.c. magnetometry experiments with sinusoidal oscillating fields ($\nu=1/T$) sampled with the corresponding Walsh sequence $w_m(t/T)$. The amplitude of the field was swept and the normalized Walsh coefficient was extracted from the measured signal $S_m(b)=\sin{(\gamma_e b\hat{f}(m)T)}$. The slope at the origin $\mu_{Vpp}=\gamma_e\hat{f}(m)T$ was compared with the value computed analytically, e.g., by choosing $\hat{f}(1)=2/\pi$ for the spin-echo sequence (m=1). The conversion factor $k=\mu_{nT}/\mu_{Vpp}~(\text{nT}\cdot\text{Vpp}^{-1})$ was calculated from the ratio between $\mu_{nT}$ calculated analytically and $\mu_{Vpp}$ measured experimentally at different frequencies. 

As shown in Supplementary Fig.~\ref{som_calibration_fig1}, the conversion factor $k(\nu)$ increases linearly in the frequency range of interest. Taking into account the intrinsic $90^{\circ}$ phase shift between the electric and magnetic fields, we have $b(\nu)=-ic\nu\Phi(\nu)$ such that $b(t)=-c\frac{d\Phi(t)}{dt}$ with $c=25.4~\mu\text{T}\cdot(\text{Vpp}\cdot\text{kHz})^{-1}$. Therefore, our coplanar waveguide behaves as the physical model of a neuron, with the magnetic field given by the first derivative of the electric field.

\subsection{Simulation of the magnetic field radiated by a single neuron}
{The creation and conduction of action potentials is the primary communication mean of the nervous system. }
{The flow of ions across neuronal membranes produce an electric field that propagates through the axon of single neurons. The electric signals carried by the action potentials radiate a magnetic field given approximately by the first derivative of the action potential~\cite{Swinney80,Woosley85}. As shown in Supplementary Fig.~\ref{som_neuronal_fig1}, we approximated the action potential by a skew normal impulse and extracted the physical parameters by fitting the simulation data obtained for a rat hippocampal mossy fiber boutons~\cite{Alle09}. The action potential was rescaled to perform proof-of-principle measurements.}

\subsection{Experimental method for Walsh reconstruction}
The Walsh reconstruction method extends the sequential phase estimation algorithm~\cite{Nielsen00b} to measuring time-varying fields with a single quantum sensor. The aim is to reconstruct an $N$-point functional approximation $b_N(t)$ to the time-varying field $b(t)\in L_1([0,T[)$ from a set of $N$ Walsh coefficients $\{\hat{b}(m)\}_{m=0}^{N-1}$.

The $m$-th Walsh coefficient is obtained by modulating the evolution of the qubit sensor with the $m$-th Walsh sequence in a Ramsey interferometry experiment. The qubit sensor is first initialized to its ground state $\ket{0}$ and then brought into a superposition of its eigenstates $(\ket{0}+\ket{1})/\sqrt{2}$ by applying a $\frac{\pi}{2}$-pulse along the $\sigma_x$ rotation axis. During the free evolution time $T$, the qubit acquires a phase difference $\phi(T)=\gamma\int_0^Tb(t)dt$, where $\gamma$ is the strength of the interaction with the external time-varying field $b(t)$ directed along the quantization axis of the qubit sensor. Under a control sequence of $m$ $\pi$-pulses applied at the zero-crossings of the $m$-th Walsh function $w_m(t/T)$, the phase difference acquired after an acquisition period~$T$ is $\phi_m(T)=\gamma\int_0^Tb(t)w_m(t/T)dt=\gamma\hat{b}(m)T$, which is proportional to the $m$-th Walsh coefficient of $b(t)$. A final $\frac{\pi}{2}$-pulse applied along the $\sigma_\theta=\cos{(\theta)}\sigma_x+\sin{(\theta)}\sigma_y$ rotation axis converts the phase difference into a measurable fluorescence signal $S_{x\theta}$.

Although performing a single measurement with $\theta=\pi/2$ is enough to extract the $m$-th Walsh coefficient, we sweep the field amplitude of $b(t)=b\,f(t)$ to better estimate it~(Supplementary Fig.~\ref{som_calibration_fig2}).  The $m$-th Walsh coefficient $\hat f(m)$ of the normalized field is obtained from the slope at the origin of the normalized fluorescence signal averaged over $M\sim10^5$ measurements: $S_m(b)=\sin{(\gamma_e b\hat{f}_mT)}=\frac{S_{x\bar{y}}-S_{xy}}{S_{x\bar{y}}+S_{xy}}\cdot\frac{S_0+S_1}{S_0-S_1}$, where $S_0$ is the fluorescence count rate of the $m_s=0$ state measured after optical polarization, and $S_1$ is the fluorescence count rate of the $m_s=1$ state measured after adiabatic inversion of the qubit with a $600$~ns frequency-modulated chirp pulse over a $250$~MHz frequency range centered around the resonance frequency.

In practical applications for which the amplitude of the field cannot be swept, the $m$-th Walsh coefficient can be equivalently measured by sweeping the phase $\theta$ of the last read-out $\frac{\pi}{2}$-pulse and fitting the normalized signal to a cosine function: $1-2S_m(\theta)=\cos{(\gamma_e\hat{b}_mT-\theta)}=\frac{S_{x\theta}-S_{0}}{S_{1}-S_{0}}$. This procedure gives an absolute estimate of $\hat{b}(m)$ rather than an estimate of the normalized coefficient $\hat{f}(m)=\hat{b}(m)/b$.

{The total measurement time for acquiring all the data of the $T=14~\upmu\text{s}$ waveform presented in Figure 4, excluding dead times associated with computer processing and interfacing, was less than $4~\text{h}$. Each of the $N=32$ Walsh coefficients were obtained from two $M^{\prime}=90$ measurements of the fluorescence signal as a function of the phase and conjugate phase of the last readout pulse to correct for common-mode noise (we note that in a well-calibrated, temperature stabilized and isolated setup this step is unnecessary). Each experimental point was averaged over $M=10^5$ repetitive measurements due to low light-collection efficiency. The length of each sequence was about $42~\upmu\text{s}$, including the two waveform measurements ($28~\upmu\text{s}$), optical polarization and readout periods ($5~\upmu\text{s}$), adiabatic inversion  ($4~\upmu\text{s}$) used for calibration purposes but not necessary, and waiting time ($5~\upmu\text{s}$).}

\subsection{Applications to neuroscience}
{In the main text we considered a paradigmatic application of the Walsh reconstruction method to monitor the dynamics of the magnetic field produced by an action potential flowing through a single neuron. Such application of the Walsh reconstruction method will be useful for neuroscience, which needs novel metrology techniques to record and manipulate neuronal activity. Indeed, the development of magnetic resonance imaging at the nanoscale will be useful for measuring transient magnetic fields produced by living cells with greater spatial and temporal resolution. These measurements will provide complementary information to conventional electrical activity recordings to better understand neurophysiology, characterize subcellular compartments, and map neural circuits. Magnetic measurements of action potentials may also detect weak local features that are not observable in electrical measurements.}

{An alternative to wide-field imaging would be to use functionalized nanodiamonds with coherence time approaching bulk diamonds as both fluorescent biomarkers and quantum probes to perform local measurements of magnetic fields and temperature. Advantages would include selective positioning, high sensitivity, and minimal invasiveness. Neuronal growth and mobility could also be studied by optically tracking the position of nanodiamonds over long time scale.}

The spatial resolution will be limited by the diffraction limit in a confocal microscopy setting (unless sub-diffraction techniques are used) and the pixel size in a wide-field imaging setup. Because neurons are living cells that move, relative displacements of the quantum sensors with respect to the living neurons will induce fluctuations in the amplitude of the magnetic field at the position of each NV center. These fluctuations will be averaged over each pixel for measurements with ensembles of NV centers, given that the pixel size is greater than the displacement of the neuron during the acquisition period.

{The Walsh reconstruction method depends on a repeatable signal assumed to be deterministic and triggered on-demand. Action potentials can be artificially created by external stimulation techniques that involve creating large potential differences across the membrane and injecting current into the system.  An initial stimulus at one end of the axon creates a potential difference across the axonal membrane; when the difference is above some threshold value, the potential suddenly spikes upwards and returns to its resting value as equilibrium is restored in the system. This spike, called the action potential, propagates along the length of the axon to the other end, where it can stimulate other connecting nerve cells. The action potential can create magnetic fields at the nT scale~\cite{Wikswo80} and each event lasts for a wide range of time scales from $1~\mu$s to $10$~ms.} 

{While there is large variability in these techniques, their key point is that the timing of events can be controlled to a high degree. Trains of hundreds action potentials can be repeatedly evoked in neuronal cells via conventional electrophysiological techniques, photo-stimulation methods, or current injection through underlying nanowire electrode arrays. The stimulation rate needs to be below the frequency threshold to avoid propagation failure~\cite{Debanne04} due to geometrical constraints, depolarization of the membrane, or hyperpolarization of the axon. The stimulation frequency threshold depends on the type of neuronal cells and range from moderate ($10-50~\text{kHz}$) to high ($200-300~\text{Hz}$), sometimes even up to $1~\text{kHz}$ for axons in the auditory pathways~\cite{Debanne04}. 
Short trains of $N$ action potentials are evoked at a stimulation rate of $100~\text{kHz}$ in $M$ stimulation-recovery cycles of length $1~\text{s}$, for a total acquisition time less than few hours, within the lifespan of neurons (several hours). Data compression methods~\cite{Magesan13} or compressed sensing techniques~\cite{Magesan13c} can be used to significantly reduce the acquisition time.}

{The fluorescence signal may be measured in a single pass  ($M=1$) in a wide-field imaging setup with large ensembles of NV centers and improved collection efficiency, e.g., with a side-collection setup. Single-pass measurement ($M=1$) of $N$ Walsh coefficients could be done by sampling a train of $N$ action potentials with the first $N$ Walsh sequences and, if needed, repeating the experiment $M$ times to improve the signal-to-noise ratio.}

\end{document}